\documentclass[12pt]{article}
\usepackage{epsfig}
\def\G{{\cal G}^{+++}}
\begin{document}
\thispagestyle{empty}
\setcounter{page}{0}
\renewcommand{\theequation}{\thesection.\arabic{equation}}

{\hfill{\tt hep-th/0402076}}

{\hfill{ULB-TH/04-02}}

\vspace{.5cm}

\begin{center} {\bf   FROM BRANE DYNAMICS TO A KAC-MOODY INVARIANT 
FORMULATION OF M-THEORIES}\footnote{Talk presented by F. Englert at the 27th
John Hopkins workshop: G\"oteborg, August 24-26, 2003. Related 
developments on the  non-linear realisation of $\G$
 are included in the text.}.

\vspace{.5cm}

Fran\c cois Englert${}^a$ and Laurent Houart${}^b$\footnote{Research  
Associate F.N.R.S.}

\footnotesize \vspace{.5 cm}

${}^a${\em Service de Physique Th\'eorique\\ Universit\'e Libre de  
Bruxelles,
  Campus Plaine, C.P.225\\Boulevard du Triomphe, B-1050 Bruxelles,  
Belgium}\\  {\tt fenglert@ulb.ac.be}

\vspace{.2cm}

${}^b${\em Service de Physique Th\'eorique et Math\'ematique and International
Solvay Institutes }\\  {\em Universit\'e Libre de Bruxelles, Campus Plaine C.P.
231}\\ {\em   Boulevard du Triomphe, B-1050 Bruxelles, Belgium}\\ {\tt
lhouart@ulb.ac.be}

\end{center}

\vspace {1cm}
\centerline{ABSTRACT}
\vspace{- 3mm}
\begin{quote}\small Theories of gravity coupled to forms and dilatons may
admit as solutions zero binding energy configurations of intersecting
closed extremal branes. In such configurations, some branes may open on
host closed branes. Properties of extremal branes reveal symmetries of the
underlying theory which are well known in M-theory but transcend
supersymmetry.  From these properties it is possible to reconstruct all
actions, comprising in particular pure gravity in $D$ dimensions,  the bosonic
effective actions of M-theory and of the bosonic string,  which upon dimensional
reduction to three dimensions are invariant under the maximally non-compact
simple simply laced Lie groups $\cal G$.  Moreover the
features of extremal branes  suggest the existence of a much larger
symmetry, namely the `very-extended' Kac-Moody algebras $\G$. This
motivates the
 construction of  explicit non-linear realisations of all simple $\G$,
which hopefully contain new degrees of freedom such as those encountered
in   string theories. They are defined without a priori reference to space-time
and are  proposed as substitutes for  original field theoretic models of gravity,
forms and dilatons. From the $\G$-invariant theories, all algebraic properties
of extremal branes are recovered from exact solutions, and  there are
indications that space-time is hidden in the infinite  symmetry structure. 
The   transformation properties of the exact solutions, which  possibly 
induce new solutions foreign to conventional theories, put into evidence the
general group-theoretical   origin of `dualities' for all $\G$. These
dualities   apparently do not require  an   underlying string theory.

\end{quote}

\newpage
\baselineskip18pt

\setcounter{equation}{0}
\addtocounter{footnote}{-1}
\section{Introduction and conclusion} Theories of gravity coupled to forms
and dilatons may admit as solutions zero binding energy configurations of
intersecting closed extremal branes \cite{allir}. In such configurations, some
branes may open on host closed branes \cite{str,tow,aehw}. Properties of
extremal branes reveal symmetries of the underlying theory, which are well known
in M-theory but transcend supersymmetry.  From these properties it is possible to reconstruct all
actions, comprising in particular pure gravity in $D$ dimensions,  the bosonic
effective actions of M-theory and of the bosonic string,  which upon dimensional
reduction to three dimensions are invariant under the maximally non-compact
simple simply laced Lie groups 
$\cal G$ \cite{ehw}.  Moreover the
features of extremal branes  suggest the existence of a much larger
symmetry, namely the `very-extended' Kac-Moody algebras $\G$. Such symmetries were first conjectured in the above mentioned
particular cases
\cite{west01,lw},  explicit representations of their Weyl group for
Kasner-type solutions were obtained for all simple $\G$ \cite{ehtw} and their
relation to the cosmological billiards  
\cite{damourhn00,damourh00,damourbhs02} was brought to light.

These facts motivate the construction of  explicit non-linear realisations
of all simple
$\G$ on cosets obtained from a modified Chevalley involution \cite{eh}.  The
$\G$-invariant  actions are  proposed as substitutes for the original field
theoretic models of gravity, forms and dilatons, and hopefully contain new
degrees of freedom such  as those encountered in string theories.  The actions
are defined on a world-line {\em a priori unrelated  to space-time}. The  
latter should then be deduced dynamically.  Such an approach to gravity and
forms, if successful,
  would  dispose of the need of   explicit diffeomorphism invariance or 
gauge invariance.  All such information should be hidden in the {\em global}
 $\G$ invariance. Although it may seem that global symmetries cannot  
contain local symmetries, in particular in view of the celebrated Elitzur
theorem
\cite{elitzur75}, it need not be the case in view of the infinite number
of generators of  
$\G$. We formulate the $\G$ invariant theory recursively from a level  
decomposition \cite{damourhn02,west02,west05}  with respect to a subalgebra
$A_{D-1}$ where
$D$  turns out to be   the space-time dimension.   Our formulation is
exploratory and does not pretend to be a final one.  No attempt is made to cope
with   fermionic degrees of freedom and we limit here our investigation  to the
classical   domain. From the $\G$-invariant theories, all algebraic properties
of extremal branes are recovered from exact solutions, and  there are
indications that space-time is hidden in the infinite level structure.  The  
transformation properties of the exact solutions, which also generate
Kaluza-Klein waves and KK-monopoles (Taub-NUT spaces) and possibly  solutions
foreign to conventional theories, put into evidence the general
group-theoretical   origin of `dualities' for all
$\G$. These dualities 
 apparently do not require  an   underlying string theory.

In Sections 2 and 3, we review  the properties of intersecting extremal
branes and of their opening, which are used in Section 4 to reconstruct
dynamically the simple simply laced Lie algebras $\cal G$ and provide
motivation for considering the very-extended Kac-Moody algebras $\G$. The
non-linear realisations of $\G$ are presented in Section 5. Extremal branes
and their duality properties are revisited in this new  framework in
Section 6.

\setcounter{equation}{0}
\section{Intersection rules for extremal branes}

We begin with a generic theory in 
$D$ dimensions  which includes gravity, one dilaton and ${\cal M}$ field
strengths  of arbitrary  form degree $p_I$  with $p_I\leq D/2$  and
arbitrary couplings to the dilaton
$a_I$.  The action is
\begin{equation} S= \frac{1}{16\pi G_N^{(D)}} \int d^D x \sqrt{-g}
\left( R - \frac{1}{2} (\partial \phi)^2 -
\sum_I \frac{1}{2 p_I !}e^{a_I \phi}F_{p_I}^2 \right)\, , \quad I=1\dots
{\cal M}\, ,
\label{action}
\end{equation} where we have not included  possible  Chern-Simons terms.  
It  can be shown that they have no effect on  the zero binding energy
solutions considered here although such terms will be important in the
following sections.

The `BPS' zero binding energy configurations of closed extremal branes
intersecting orthogonally  are obtained by first specialising to metrics of
the following diagonal form,
\begin{equation} ds^2=-e^{2p^{(1)}} d\tau^2+\sum_{\mu=2}^{D-p}  
e^{2p^{(\mu)}}
(dx^\mu)^2+\sum_{\lambda=1}^pe^{2p^{(D-p+\lambda)}}(dy^\lambda)^2\, ,
  \label{metric}
\end{equation} where $y^\lambda$  label  $p$  compact coordinates.   The
functions $p^{(\alpha)}~ (\alpha=1,2,\dots D)$ depend only on the   
transverse coordinates $x^\mu$ in the non-compact dimensions and  allow
for   multi-centre solutions. We choose $p$ so that all branes are wrapped
in the compact dimensions and that no compact dimension is transverse to
all branes. Thus for each brane  $p\ge q_A$ where $q_A$ is the dimension of
the   brane. If
$q_A<p$, we take  a lattice of  $q_A$-branes in the  compact directions  
transverse to the brane and average over them. Here $q_A$ designates either
an   electrically charged $q_A^e$-brane  with respect to a 
$p_I$-form field strength $F_{p_I}$, or its  dual $q_A^m$-magnetic brane.
For $\cal N$ intersecting branes, we write
\begin{equation}
\label{many} p^{(\alpha)}= \sum_{A=1}^{\cal N} p^{(\alpha)}_A\quad \hbox {and}
\quad
\phi=
\sum_{A=1}^{\cal N} \phi_A
\end{equation} with
\begin{equation}
\label{embedding} (p+3-D)\,   p^{(\mu)}_A= p^{(1)}_A+ \sum_{\lambda=1}^p  
p^{(D-p+
\lambda)}_A\qquad
\mu=2,\dots, D-p\, .
\end{equation} Eq.(\ref{embedding}) may be interpreted as the extremality
condition of the component branes \cite{aeh}.  For electric and magnetic brane
potentials, we take
\begin{eqnarray}
\mbox{Electric}& :& A_{\tau
\lambda_1\dots\lambda_{q_A^e}}=\epsilon_{\tau
\lambda_1\dots\lambda_{q_A^e}}E_A^e(\{x^\nu\})\, ,
\label{electric}\\
\mbox{Magnetic}& :& \widetilde{A}_{\tau
\lambda_1\dots\lambda_{q_A^m}}=\epsilon_{\tau
\lambda_1\dots\lambda_{q_A^m}}
  E_A^m(\{x^\nu\})\, ,
\label{magnetic}
\end{eqnarray} where $ \widetilde{A}$ is the (magnetic) potential of   the
dual field strength
$\widetilde{F}$  defined by
\begin{equation}
\label{hodge}
\sqrt{-g} e^{a_I\phi}F^{\mu_1 \dots \mu_{p_I}}={1\over (D-p_I)!}
\epsilon^{\mu_1 \dots \mu_{p_I}\,  \nu_1 \dots \nu_{D-{p_I}}}
\widetilde{F}_{\nu_1 \dots \nu_{D-{p_I}}}\, .
\end{equation}

 Using these ans\"atze, the Einstein equations and the equations for the
dilaton and the forms yield
\begin{eqnarray}
  p^{(1)}_A &=& - \frac{D-q_A-3}{\Delta}  \ln H_A\, ,
\nonumber\\
  p^{(\mu)}_A &= &  \frac{q_A+1}{\Delta} \ln H_A\, ,
\nonumber\\ p^{(D-p+\lambda)}_A&=& 
\frac{\delta^{(\lambda)}_A}{\Delta}\ln H_A\, , \nonumber\\
\phi_A &=& \frac{D-2}{\Delta}\varepsilon_A \, a \ln H_A\, ,
\label{phicond}
\end{eqnarray} where $H_A(\{x^\nu\})$ is a harmonic function related to  
the
$E_A$ by $H_A=
\sqrt{2(D-2)/\Delta}\, E_A^{-1}$ and
\begin{equation}
\label{delta}
\Delta= (q_A+1)(D-q_A-3)+\frac{1}{2}a_A^2(D-2)\, .
\end{equation}
  In Eq.(\ref{phicond})
$\delta^{(\lambda)}_A=-(D-q_A-3)$ or $(q_A+1)$ depending on wether
$y^\lambda$ is parallel or perpendicular to the
$q_A$-brane. The factor $\varepsilon_A$ is
$+1$  for  an electric brane and $-1$ for a magnetic one.  The harmonic  
functions
$H_A(\{x^\nu\})$ allow for parallel branes and are given  (for   
$D-p>3$) by
\begin{equation}
\label{branes} H_A=1+\sum_k \frac{Q_k}{|x^\mu-x^\mu_k|^{D-p-3}}\, ,  
\label{multicenter}
\end{equation} where the $x^\mu_k$ label the positions in non-compact
space-time of the branes with charge
$Q_k$.  For ${\cal N}>1$, Eqs.(\ref {phicond})  are  restricted by 
algebraic conditions stemming from the equation for the non diagonal
components of the curvature tensor. These are, for each  pair
$(A,B)$ of distinct 
$q$-branes of dimensions $(q_A, q_B)$,       the number of dimensions $\bar
q \, (-1\leq {\bar q} \leq q_A,q_B$) on which they intersect\footnote{
$\bar{q}$ must be integer and the case
$\bar{q}=-1$ is  relevant. It can be interpreted in  terms of instantons in
the Euclidean, in which case the time coordinate   need not be longitudinal
to all  branes.} in terms of the total number of space-time dimensions $D$
and of the field strength couplings to the dilaton. These {\em intersection
rules} read \cite{aeh}
\begin{equation}
\bar{q}+1=\frac{(q_A+1)(q_B+1)}{D-2}-\frac{1}{2}
\varepsilon_A a_A \varepsilon_B a_B\, . \label{intrule}
\end{equation}

Note that  such zero energy binding configurations were originally
considered in the context of M-theory `phases' but may arise classically
without  supersymmetry.

\setcounter{equation}{0}
\section{The opening of branes}

We  now  analyse the  breaking of closed extremal branes into open branes
terminating on closed ones. We  consider the above BPS configurations  in
the special case when
${\bar q}$ has the  same dimension as the potential boundary of one of the
two constituent branes, i.e
$q_A-1$ or
$q_B-1$, and study its possible opening.  Such opening requires  the
addition of Chern-Simons terms to the action Eq.(\ref{action}) and may
enlarge the brane content of the theory. We shall see in Section 4 that,
under some conditions, such openings  fully determine   the theory and 
relate brane dynamics to the existence of a symmetry. The presentation
given in this section is a generalisation of the one performed in the
context of M-theory \cite{str,tow,aehw}.

Let us review  how extended objects  carrying a conserved charge can be
opened. The main obstacle towards  opening of branes is charge
conservation. Generically,  the charge density of a
$q$-brane is measured by performing an integral of the relevant field
strength on a
$(D-q-2)$-dimensional sphere $S^{D-q-2}$ surrounding the brane in its
transverse space,
\begin{equation}  Q_q \propto \int_{S^{D-q-2}} *F_{q+2}\, .
\label{charge}
\end{equation}
 If the brane is open, we can slide the $S^{D-q-2}$ off a loose end and
shrink it to zero size. This would imply the  vanishing of the charge and
hence a violation of charge conservation. This conclusion is avoided if, in
the above process, the
$S^{D-q-2}$  necessarily goes through a region in which the equation
\begin{equation}  d*F_{q+2}=0
\label{fmot}
\end{equation} no longer holds.  This is  the case when the open brane ends
on some other  one.

In the framework of M-theory  the source terms needed in Eq.(\ref{fmot}) 
to ensure charge conservation for the open branes originate from two
requirements whose interplay  leads to a consistent picture.  On the one
hand there are space-time Chern-Simons type terms in supergravity which
allow for charge conservation for well defined pairing of open and `host'
branes
\cite{tow}. On the other hand   the world-volume effective actions
\cite{str} for the branes of M-theory relate world-volume fields and 
pullbacks of space-time fields, and  gauge invariance \cite{wit} for open
branes ending on the `host' brane implies that the end of the open branes
acts as a source for the world-volume field living on the closed `host'
brane. 

In reference \cite{aehw} a systematic study in M-theory of all the zero
binding energy configurations Eq.(\ref{intrule}) corresponding to 
${\bar q} = q_A-1$ (with $q_A \leq q_B$) was performed. It was shown that
in all cases it was possible to open the $q_A$-brane along its intersection
with the $q_B$-brane. The crucial ingredient was the presence of the 
appropriate Chern-Simons terms in the supergravity Lagrangians for each
case.

Here we propose to reverse the logic. Starting with a Lagrangian of type
Eq.(\ref{action}) and having zero binding energy configurations between
branes we will ask that, if
${\bar q}=q_A-1$,   the corresponding $q_A$-brane  open  on the
$q_B$-brane.  This will {\it determine} the form of the Chern-Simons terms
one has to add to Eq.(\ref{action}) and will also  in some cases require
the {\it introduction of new field strength  forms} 
$F_{n_I}$. One then proceeds iteratively. 

We illustrate the role of the Chern-Simons term (see also \cite{tow,
aehw})  by taking as example    a theory  with only one $n$-form 
$F_{q^e+2}$ and dilaton coupling in Eq.(\ref{action})  such that  the
intersection rule between the electric $q^e$-brane and the  magnetic
$q^m$-brane ($q^m=D-q^e-4$) is ${\bar q}=q^e-1$.

We  modify the Eq.(\ref{fmot}) for $F_{q^e+2}$ in order to be able to allow
the opening of $q^e$ on $q^m$ by the addition   to the action
Eq.(\ref{action}) of the Chern-Simons term
\begin{equation}\label{cs}
\int A_{q^e+1}\wedge F_{q^e+2}\wedge F_{D-2q^e-3}\, ,
\end{equation} and of a  standard   kinetic energy term for the 
new\footnote{ If $D=3q^e+5$ there is no need to introduce a new field
strength as the Chern-Simons terms can be build with the field strength
$F_{q^e+2}$.}  field strength  $F_{D-2q^e-3}$. Its dilaton coupling is
chosen such that the intersection rules Eq.(\ref{intrule}) give  integer 
intersection dimension between the new extremal electric
$(D-2q^e-5)$-brane and its dual magnetic
$(2q^e+1)$-brane.   Charge conservation now reads
\begin{equation}  d*F_{q^e+2} = F_{q^e+2} \wedge F_{D-2q^e-3} + Q^e
\delta_{D-q^e-1}\, . \label{emcase}
\end{equation} Here  wedge products are defined up to
signs and  numerical factors. In the r.h.s. of Eq.(\ref{emcase}) the first
term comes from the  variation of the   Chern-Simons term and  the second
one is the $q^e$-brane charge density. 
$\delta_{D-q^e-1}$ is the Dirac delta function in the directions transverse
to the
$q^e$-brane. We  introduce here an explicit source term for the electric
brane since, to study its opening, we want to extend to the branes
themselves the validity of the usual closed brane solution .  Such term is
required because the equations of motion from which the intersecting brane
solutions were derived do not contain any source term and are therefore
valid only outside the sources.

It is straightforward to show that Eq.(\ref{emcase}) renders opening  
consistent with charge conservation and that  new branes arising from
$F_{D-2q^e-3}$ have to be added. From Eqs.(\ref{cs}) and  (\ref{emcase}),
the addition of such branes can be determined in a picturesque way.
Namely,  when a brane opens on a closed `host' brane,  the boundary appears
from the world-volume point of view  as a charged object  under a
world-volume field strength living  in the closed brane. The world-volume
Hodge dual of this object is   the boundary of an other brane which can
also be consistently opened on  the same closed  `host'  brane. The field
strength associated to this new brane is precisely the one appearing in the
Chern-Simons ensuring the consistency of the opening of the  brane we
started with. 

In brief, having  a theory  of type Eq.(\ref{action})  in which some zero
binding energy configurations  give potential boundaries, it is possible to
complete it in a well-defined way by adding Chern-Simons terms\footnote{The
coefficients of the Chern-Simons terms Eq.(\ref{cs}) are not fixed in this
qualitative discussion. Their precise values are  important when one will
discuss the potential symmetries.  In the framework of M-theory they are
usually fixed by supersymmetry.  Nevertheless it is possible to fix them or
at least quantise them generically using only consistency arguments without
appealing to supersymmetry, see for instance \cite{bachas, bg}.},  and when
necessary new form field strengths (hence new branes) in order to ensure
consistency of brane  opening with charge conservation. In section 4, such
dynamical requirement will be at work to  reconstruct purely from brane
considerations theories whose dimensional reduction is characterised by a
coset symmetry.

\setcounter{equation}{0}
\section{From brane dynamics to $\cal G$ and $\G$ }
\subsection{Intersection rules and dimensional reduction}

We  perform the dimensional reduction to three dimensions of  a  generic
theory described by an action  Eq.(\ref{action}), possibly with the
addition of  Chern-Simons terms and discuss the possible emergence of a
coset symmetry.   We  relate the onset  of  such symmetry to the
intersection  rules for BPS configurations between  the extremal branes of
the theory considered.

Starting with the theory defined in $D$ dimensions, we  compactify  to
$D-1$  dimensions while  remaining  in the Einstein frame with  the
standard 1/2  kinetic term normalisation for the new scalar.   We use the
notation $\phi_2$ for the scalar appearing in the first step of the
dimensional reduction and  rename 
$\phi_1$   the dilaton $\phi$   already present in the uncompactified theory
defined by Eq.(\ref{action}). The compactified coordinate is $x^{D-1}$, the
uncompactified coordinates are $x^\mu$ where
$\mu=0 \dots D-2$, and
\begin{equation}
\beta_D = \sqrt{1\over2(D-1)(D-2)}\, .
\label{alpha}
\end{equation} The gravitation part of the action Eq.(\ref{action}) becomes
\begin{eqnarray}
\int d^D x \sqrt{- g_D} \,  R_D &=& \int  d^{D-1} x  
\sqrt{-g_{D-1}}\, (\,  R_{D-1} - {1 \over 2}
\partial_\mu\phi_2\partial^\mu\phi_2
\nonumber \\ &-& {1\over 4}e^{-2(D-2) \beta_{D-1}\phi_2}\,
F_{\mu\nu}F^{\mu\nu}\, )\, .
\label{gredu}
\end{eqnarray} where $ F_{\mu\nu}=\partial_\mu A_\nu - \partial_\nu A_\mu$.

For each $n_I$-form field strength
$F_{n_I}$ in Eq.(\ref{action}) we get after reduction
\begin{eqnarray}
\int d^{D} x\, \sqrt{-g_D}\, {1 \over 2 n_I!} \, e^{a_I  
\phi_1} F_{n_I}^2  &=& \int d^{D-1} x  \sqrt{-g_{D-1}}\, ( \, {1 \over 2
n_I!}\, e^{a_I 
\phi_1-2(n_I-1)\beta_{D-1}\phi_2}\,  F_{n_I}^{{\prime}2} \nonumber \\ 
&+&   {1\over 2 (n_I-1)!}\,  e^{a_I  \phi_1 + 2(D-1-n_I) \beta_{D-1}
\phi_2}\,  F_{n_I-1}^2 \, )\, ,
\label{freduc}
\end{eqnarray} where 
\begin{equation}
 F_{\mu_1 \dots \mu_n}^{\prime}= F_{\mu_1 \dots \mu_n}-
 n\, F_{ [ \mu_1 \dots \mu_{n-1}} A_{\mu_n ] }\, .
\label{fmod}
\end{equation}

We can repeat this procedure step by step to obtain the theory on a
$p$-torus. One has then obviously $p$ scalars $\phi_j$ with 
$j=2 \dots p+1$  parametrising the radii of the torus,   coming from  the
diagonal components of the metric in the compact dimensions. Additional
scalars denoted
$\chi_{\vec\alpha}$ arise  from several origins. They come from  potentials
$A^k_\mu$   which arise when reducing gravity from $D+1-k$ to $D-k$  and also
from  the potentials associated to the $F_{n_I}$ (when $p \geq n_I-1$) with
indices in the compact dimensions. In addition the $n$-form field strengths give
additional scalars when
$p=D-n-1$ by dualising them. In particular when we reach $D=3$ the 
$F^k_{\mu \nu}$  (with
$k=1
\dots D-3$) coming from the gravity part of the action (i.e. the
graviphotons) can be dualised to scalars, and we are left with only 
scalars. The action takes then the  form
\begin{equation} S = \int d^3 x \, \sqrt{-g_3} \, 
\left(R_3 - {1 \over 2} \partial_\mu \vec{\phi}\cdot\partial^\mu \vec \phi
- {1\over2}
\sum_{\vec
\alpha}e^{\sqrt{2} \vec\alpha \cdot \vec\phi}
\partial_\mu\chi_{\vec\alpha}\partial^\mu\chi_{\vec\alpha}+\dots\right)\, ,
\label{clag}
\end{equation} where $\vec\phi = (\phi_{D-2},...,\phi_1)$, the $\vec\alpha$
are  constant ($D-2$)-vectors\footnote{ The normalisation factor
$\sqrt{2}$ has been chosen for convenience. It will eventually correspond
in the simply laced case to the standard normalisation of the roots, namely 
$\vec\alpha \cdot \vec \alpha =2$.} characterising  each
$\chi_{\vec\alpha}$. If we start in
$D$ dimensions without dilaton the vectors are of course
($D-3$)-dimensional as $\phi_1$ is absent in that  case. The ellipsis in
Eq.(\ref{clag}) stands for terms of order higher than quadratic in the 
$\chi_{\vec\alpha}$ scalars. They come from the modification of the field
strengths    Eq.(\ref{fmod}) in the dimensional reduction process and also
from 
  possible Chern-Simons terms   in the uncompactified theory.

The action Eq.(\ref{action})  dimensionally reduced to three dimensions 
has a ${\cal G}/{\cal H}$  symmetry {\it if} the vectors $\vec\alpha$
obtained from the compactification can be identified with the positive
roots of a group ${\cal G}$ and {\it if}, when necessary,  some precise
Chern-Simons terms are added in the uncompactified  theory \cite{pope}.  The
requirement that the 
$\vec\alpha$ correspond to positive roots is thus a necessary condition to
uncover a symmetry \cite{lw}.

Recall  first the well-known  dimensional reduction  of pure gravity (see
Eq.(\ref{gredu})) down to three dimensions, which  leads to ${\cal G}=
SL(D-2)$ whose algebra is
$A_{D-3}$.  The scalars corresponding to the simple roots of $A_{D-3}$ are
of two kinds. 

There are first $D-4$ scalars which are the components $A^k_{D-k-1}$ of the
potentials  coming  from
$g_{D-k,D-k-1}, \ k=1 \dots D-4$.   These are obtained by performing the
``fastest'' reduction  on the potentials $A_\mu^k$ (see Eq.(\ref{gredu}))
to obtain  a scalar going from
$D-k$ to $D-k-1$ when compactifying on $T^{k+1}$.  The corresponding simple
roots 
$\vec{\alpha}^g_k$ are given by
\begin{eqnarray}
\vec\alpha^g_k &=& \sqrt{2}(\underbrace{0, \dots ,0}_{D-4-k\,
\rm{terms}},(D-k-3)\beta_{D-k-1},
 - (D-k-1)\beta_{D-k}, 
\underbrace{0, \dots , 0}_{k-1 \,   {\rm terms}}  ;0)\, ,
\nonumber \\ k&=& 1 \dots D-4\, .
\label{gchain}
\end{eqnarray}

They define a subalgebra  $A_{D-4}$. We have indeed
\begin{equation}
\vec\alpha^g_k\cdot\vec\alpha^g_l = \left\{\matrix{ 2\cr -1\cr 0\cr} \ 
\matrix{k=l\cr|k-l|=1\cr|k-l|\ge2\cr}\right.
\label{scalarp}
\end{equation} Reading from right to left, the first component of
$\vec\alpha^g_k$  associated to the dilaton
$\phi$ in the original uncompactified theory Eq.(\ref{action})  is always
zero. The corresponding Dynkin diagram with $D-4$ nodes, which  from now
on  we will refer to as  {\it the gravity line}, is depicted in Fig.1.
\vskip1cm
\hskip 2.5cm
\epsfbox{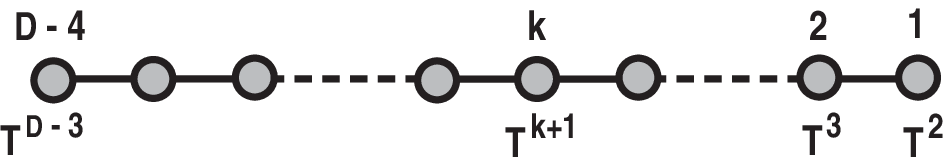}
\vskip.5cm

\begin{quote}
\begin{center}
\baselineskip 12pt {\small Fig.1.  The gravity line.

 Dynkin diagram of $A_{D-4}$ generated by the dimensional reduction to 3
dimensions.}
\end{center}
\end{quote} The remaining scalar corresponding to the missing simple root
leading to  the full
$A_{D-3}$ comes from dualising in three dimensions  the first vector
(graviphoton) that arises in the stepwise procedure namely the vector
appearing already in $D-1$ dimensions. The corresponding simple root is
\begin{equation}
\vec\alpha^{gp}= \sqrt{2} (\beta_3, 
 \dots ,\beta_{D-3},\beta_{D-2},(D-2) \beta_{D-1};0)\, .
\label{gravip}
\end{equation} Note  that this simple root $\vec\alpha^{gp}$ has a 
non-vanishing scalar product with $\vec\alpha^g_1$ (i.e with the simple
root already appearing when compactifying on $T^2$).  One has indeed 
$\vec\alpha^g_k \cdot \vec\alpha^{gp} = - \delta_{k,1}$.  Consequently it
attaches itself to the right of the gravity line.  The  other  ${1 \over 2}
(D-4) (D-3)$ scalars coming from the reduction of gravity down to three
dimensions give  all the positive roots of $A_{D-3}$.

We now turn to theories with forms given by Eq.(\ref{action}) and  consider
a $n_{A}$-form $F_{n_A}$ with dilaton coupling $a_A$ (and $n_A \leq D/2$).
Let us consider the first scalar arising  from the $n_A$-form upon
dimensional reduction up to $p=n_A-1$. The vector 
 $\vec\alpha_{n_A}^e$ associated to this scalar\footnote{The other scalars
obtained by further  dimensional reduction  give
$\vec\alpha$-vectors that are linear combinations with positive integer
coefficients of
$\vec\alpha_{n_A}^e$ and of the
 $\vec\alpha^g_k$.} will from now on be  called  {\it the would-be electric
root}. It is given by
\begin{equation}
\vec\alpha_{n_A}^e=   (\underbrace{0, \dots ,0}_{D-n_A-2 \, {\rm terms}},
\underbrace{b_{(n_A,D)}\, \beta_{D-n_A+1},  \dots ,\, b_{(n_A,D)}\, 
\beta_{D-2}, b_{(n_A,D)}\, 
\beta_{D-1}}_{n_A-1 \, {\rm terms}} ;{a_A \over \sqrt{2}})\, ,
\label{eroot}
\end{equation}

with
\begin{equation} b_{(n_A,D)}= \sqrt{2} \, (D-n_A-1)\, .
\label{bdef}
\end{equation}First, we compute the scalar product of the would-be electric
root with the gravity line\footnote{ The scalar product of the  would-be
electric root with the graviphoton is one for any $D$ and $a_A$. This
implies that when forms are present in a theory with symmetry, the
graviphoton is  never a simple root. Consequently we  focus on the gravity
line.}.  Using Eq.(\ref{gchain}) and Eq.(\ref{eroot}) we  find, {\it for
any D and dilaton coupling} $a_A$,
\begin{equation}
\vec\alpha^g_k\cdot\vec\alpha_{n_A}^e = -\delta_{k,n_A-1}\, .
\label{spg}
\end{equation}
\hskip 3.5cm
\epsfbox{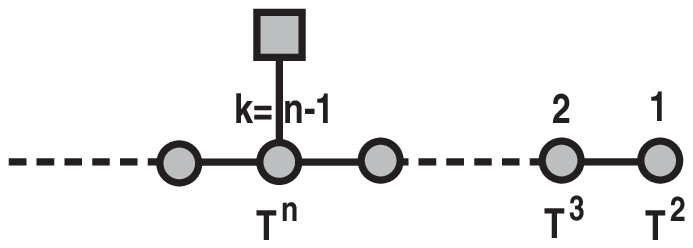}
\vskip .5cm
\begin{quote}\begin{center}
\baselineskip 12pt
 {\small Fig.2. Would-be Dynkin diagram.

 The would-be electric root associated to  $F_{n}$  is represented by a
shaded square.}
\end{center} 
\end{quote}
 Using the scalar product  Eq.(\ref{spg}) of the would-be electric root with
the gravity line,  we can draw a would-be Dynkin diagram  where the
would-be electric root associated to  $F_{n_A}$  is connected to the
$(n_A-1)^{th}$  node of the gravity line as depicted in Fig.2.

We then evaluate the length of the would-be electric root using 
Eq.(\ref{eroot}). We find
\begin{eqnarray} 
\vec\alpha_{n_A}^e \cdot\vec\alpha_{n_A}^e &=&  {(n_A-1)(D-n_A-1) \over
(D-2)} + {a_A^2
\over 2}\, , \nonumber \\ &=& {(q^e_A+1)(q^m_A+1) \over (D-2)} + {a_A^2
\over 2}\, ,
\label{irredu}\\ &=& {\bar q}^{(e_A,m_A)} +1\, . \nonumber
\end{eqnarray} We thus see that the square length of the would-be electric
root associated to
$F_{n_A}$ can be written in terms of the intersection rule  equation
Eq.(\ref{intrule}) giving the intersection between the electric and the
magnetic brane charged under $F_{n_A}$.

From now on we will restrict ourselves to simply laced groups. In  our
normalisation  all their roots are of square length two.  In order for the
would-be root to be a root, one must  have
$\vec\alpha_{n_A}^e \cdot\vec\alpha_{n_A}^e=2$. Consequently   {\it the
existence of a BPS configuration in the original theory, consisting of an
electric extremal
$p$-brane ($p \geq 1$) and its magnetic dual whose intersection is ${\bar
q}^{(e_A,m_A)}=1$,  is   a necessary condition to have after dimensional
reduction an enhanced simply laced Lie group symmetry.} 

\subsection{Dynamical reconstruction of Lie groups} 

Our starting point is a theory given by Eq.(\ref{action}) in $D$ 
dimensions  with only one
$n_A$-form field strength $F_{n_A}$ and its dilaton coupling $a_A$. We
will  fix the dilaton coupling such that there exists a zero-binding energy
configuration between the electric
$q^e_A$-brane ($q^e_A= n_A-2$)  and the magnetic $q^m_A$-brane with ${\bar
q}^{(e_A,m_A)}=1$. As explained in the previous section this is a necessary
condition in order to find  a new symmetry. Once the dilaton coupling of
the  form is fixed, we  require that, when the dimensionality of an
intersection permits opening,
 the latter is consistent with charge conservation.  Namely we  impose
that, if ${\bar q}^{(e_A,m_A)}= q^e_A-1$,  the electric brane  open  on the
magnetic brane.   As explained in Section 3, this  requires the
introduction of  a  specific Chern-Simons term in  the action, which may
contain a  new form field strength
$F_{n_B}$. The dilaton coupling $a_B$ of the new form field strength is
then again fixed modulo its sign by the necessary condition 
${\bar q}^{(e_B,m_B)}=1$.  The intersection rules between the extremal branes
corresponding to the different forms can be calculated and this fixes the
relative signs of the dilaton couplings. We can then check if new openings
are possible. If it is the case, we  iterate the procedure until 
consistency of all the  openings are ensured. In this way, we are able
to  reconstruct all the maximally oxidised theories corresponding to the
simply laced groups
$\cal G$ \cite{ehw}, that is all actions Eq.(\ref{action}) which lead upon
dimensional reduction to such
 group $\cal G$ and which are not dimensional reduction from an  action in
higher dimension \cite{max}. This leads  to the following conclusion: {\it The
existence of BPS configurations  with}
${\bar q}=1$ {\it between any electric extremal $p$-brane $(p \geq 1)$ and
its magnetic dual, along with the  requirement of consistency of brane
opening in the
 original uncompactified theory  (characterised by at most one dilaton),
is  a necessary and sufficient condition to have a theory whose dimensional
reduction  down to three dimensions has a simple simply laced group ${\cal
G}$ symmetry}.

Using Eq.(\ref{irredu}) the condition ${\bar q}^{(e_B,m_B)}=1$ yields  up
to a sign the dilaton coupling
\begin{equation} a_n^2(D) = 2 \frac{(1-q) D + (q^2+4q-1)}{D-2}\, .
\label{dilcon}
\end{equation}  with
\begin{equation}
\label{condi} a_n^2 \geq 0\, .
\end{equation}  The procedure outlined above gives the following relations
between $q, D$, the Chern-Simons terms, and the Lie group symmetry of the
theory reduced to three dimensions exhibited in Table I
\begin{center}
\begin{tabular}{||c||c||c||l||l||c||}
\hline $q$&$D$&$a^2_{q+2}$&~ Openings&$\qquad ~\, ÊC.S.$&${\cal G}$\\
\hline\hline
$0$&$-$&$ 2(D-1)/( D-2) $&$-$&$-$&$A_{D-1}$ \\
\hline
$1$&$-$&$8/( D-2)$&$-$&$-$&$D_{D-2}$ \\
\hline
$2$& $11$&$ 0$&$2\cap 5=1$&$A_3\wedge F_4\wedge F_4  $&$E_8$\\
\hline
$2$& $10$&$1/4$&$2\cap 4=1,\dots $&$A_3 \wedge F_4\wedge
H_3,\dots $&$E_8$\\
\hline
$2$&$9$&$4/7$&$2\cap 3=1 $&$A_3\wedge F_4\wedge F_2 $&$E_7$\\
\hline
$2$&$8$&$1$&$2\cap \bar 2=1 $&$A_3\wedge F_4\wedge F_1 $&$E_6$\\
\hline
$3$&$10$&$0$&$-$&$-$&$E_7$\\
\hline
\end{tabular}

{\small   Table I : Reconstruction from branes of  the simple simply laced
Lie groups  
$G$ .}
\end{center}

We see indeed that all simple simply laced Lie groups (in their maximally
non compact form) are recovered from the intersecting extremal brane
solutions and their openings from actions of the type Eq.(\ref{action})
with Chern-Simons terms. The action which upon dimensional reduction to
three dimensions exhibit these Lie group symmetry are fully determined.

\subsection{From $\cal G$ to $\G$}

Taking into account the intersection rules Eq.(\ref{intrule}),  one
verifies that Eq.(\ref{phicond}) implies  
\begin{equation}
\label{gplus}
\sum_{\alpha=1}^D (  p^{(\alpha)})^2 -\frac{1}{2}(\sum_{\alpha=1}^D  
p^{(\alpha)})^2 +
\frac{1}{2}
\phi^2=\frac{D-2}{\Delta}\sum_A \ln^2 H_A(x^a)\, ,
\end{equation}  and we recall that the extremal branes in the intersecting
brane configurations satisfy the extremality conditions
Eq.(\ref{embedding}). Eqs. (\ref{embedding}) and (\ref{gplus}) indicate that
the  solutions Eq.(\ref{phicond}) carry a group-theoretical significance in
the triple (or very-) Kac-Moody extension $\G$ of $\cal G$, for all simple
groups $\G$ (simply laced or not)  as we  now show.

We first recall how all simple  Lie algebra
$\cal G$ can be embedded in a very-extended Kac-Moody algebra $\G$. The
simple roots of ${\cal G}^{+++}$ are given by adding  two nodes to the
gravity line of  the Dynkin diagram of the affine extension $\cal G^+$ of
$\cal G$, thus increasing by three the rank of 
 $\cal G$ \cite{ogw}.  The resulting Dynkin diagrams for $\G$ are shown in
Fig.3. As in the case of Lie algebras, the Dynkin diagrams yield the Cartan
matrix of
$\G$ which is then entirely determined from the Serre relations. 

The group
$SL(D)$ defined by this triple extended gravity line can be extended to
the  full deformation group $GL(D)$ whose algebra, generated by  $D^2$
generators
$K^a_{~b}~,~a, b=1,\dots, D$ , is a subalgebra of  
${\cal G}^{+++}$. The $K^a_{~b}$ satisfy the following commutation relations
\begin{equation}
\label{Kcom} [K^a_{~b},K^c_{~d}]
=\delta^c_{~b}K^a_{~d}-\delta^a_{~d}K^c_{~b}\,  .
\end{equation} One considers the Cartan subalgebra of ${\cal G}^{+++}$
generated by the
$K^a_{~a}$ and $s= r-D$ abelian generators $R_u$ where $r$ is the rank of
${\cal G}^{+++}$. We write the corresponding abelian group element ${\cal
V}_{\rm abelian}$ as
\begin{equation}
\label{explicit} {\cal
V}_{\rm abelian}=\exp (-\sum_{a=1}^D p^{(a)} K^a{}_a-\sum_{u=1}^s
\phi^u R_u)\, .
\end{equation}  and tentatively identify $p^{(a)}$ and $\phi^u$ in
Eq.(\ref{explicit}) with (the log of) the diagonal vielbein in a triangular gauge\footnote{We can
indifferently label the $p^{(a)}$ by a curved or a flat index, as it is
uniquely defined by the   diagonal vielbein in the triangular gauge. The
  position of this index as a subscript or superscript is  a matter   of
convention and has no tensor significance.}
$e^a_\alpha(x) ,\alpha\leq a $ and $\phi^u$ with dilaton fields $\phi^u(x)$
in any action $\widetilde S$ describing  a maximally oxidised theory\footnote{A
geometrical motivation of this identification is given in \cite{ehtw} and an
alternate motivation has been proposed in \cite{west00} by extending $GL(D) $ to
$IGL(D)$. In this work the justification of the  identification
follows from the analysis of Section 5. }. The actions
$\widetilde S$ generalise to all simple Lie groups the ones reconstructed in
Section 3 for simply laced groups. It takes the form
\begin{equation}
\label{oxid}
  \widetilde S = \frac{1}{16\pi G_N^{(D)}} \int d^D x \sqrt{-g}
\left[ R - \frac{1}{2} \sum_{u=1}^s (\partial \phi^u)^2 -
\frac{1}{2}\sum_I \frac{1}{ p_I !}\exp(\sum_{u=1}^s a_I^u  
\phi^u)F_{p_I}^2 \right] + C.S.\, ,
\end{equation} where $C.S.$ represents   Chern-Simons terms that are  
required for some groups $\cal G$.

\vskip .3cm
\hskip .4cm\epsfbox{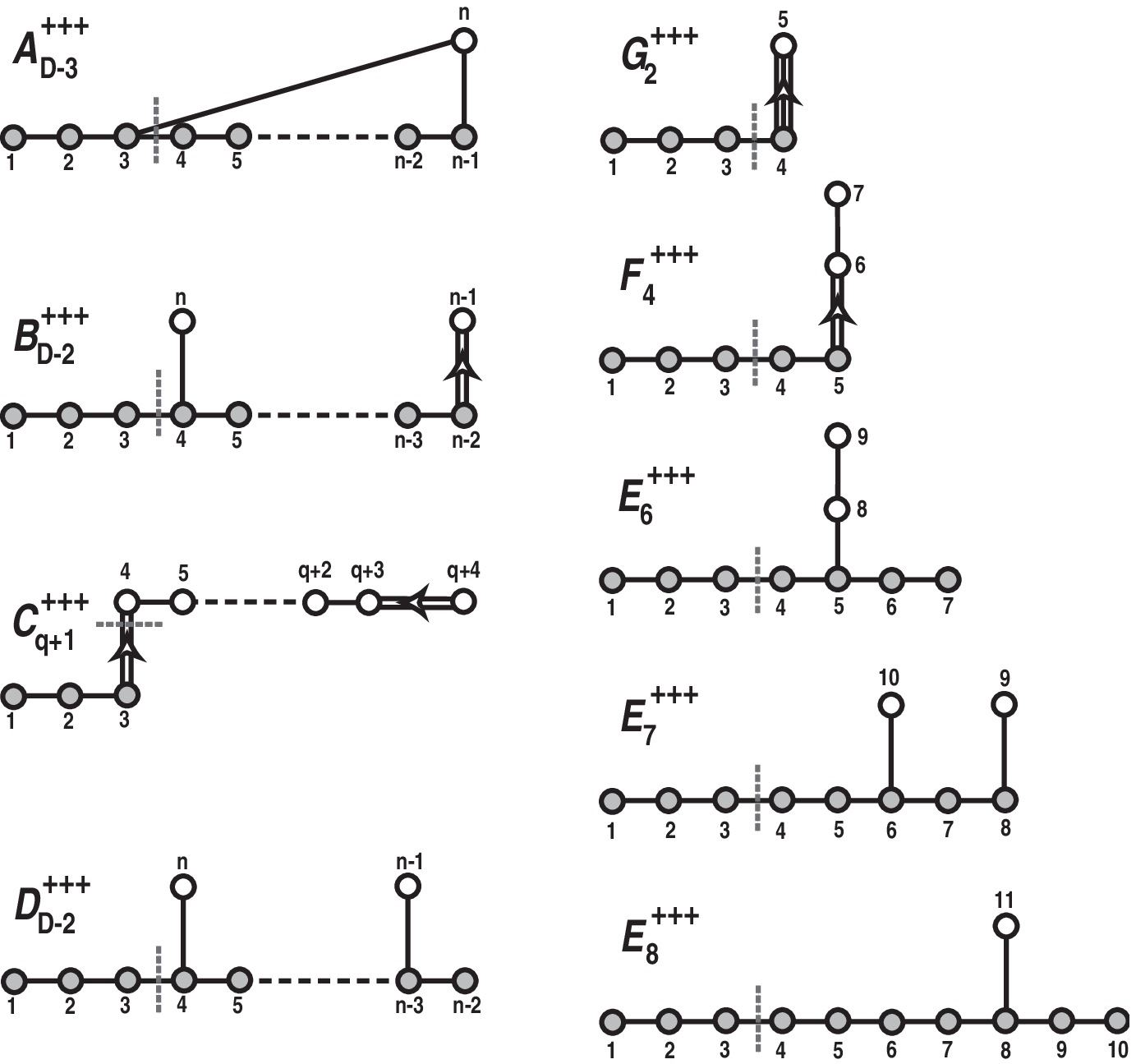}
\begin{quote}\begin{center}
\baselineskip 10pt {\small Fig.3.  Dynkin diagram of
$\G$.}
\end{center} {\small The nodes of the gravity line are shaded. The   Dynkin
diagram of $\cal G$ is that part of the diagram of $\G$  which sits on  
the right of the dashed line. The first three nodes define the Kac-Moody
extensions.}
\end{quote}

Given any symmetrisable  Kac-Moody algebra there  exists, up to a numerical
factor, a   unique scalar product defined on the algebra that is invariant
under the   adjoint action of the algebra \cite{kac83}.   For a
finite-dimensional simple Lie algebra   this is just the Killing form which
can be expressed as the trace of the   generators in any finite-dimensional
representation.  In the Cartan subalgebra described in the $L_i =\{
K^a{}_a, R_u\} $ basis it is \cite{ehtw}
\begin{equation}
\label{cartanscalar}
\langle K^a{}_a, K^b{}_b\rangle= \delta_{ab} -\frac{1}{2}{\bf
\Xi}_{ab}\quad;\quad \langle R_u, R_v\rangle=\frac{1}{2}\delta_{uv}\quad
;\quad
\langle  K^a{}_a, R_v\rangle= 0\, , 
\end{equation} where ${\bf \Xi} $ is the matrix with all entries equal to
one. The matrix ${\bf G}_{ij} =\langle L_i,L_j\rangle$ is invariant under
the group of Weyl transformations (and of outer automorphisms) of $\G$.
Hence the quadratic form
\begin{equation}
\label{cartanquad}
\sum_{\alpha=1}^D (  p^{(\alpha)})^2 -\frac{1}{2}(\sum_{\alpha=1}^D  
p^{(\alpha)})^2 +
\frac{1}{2}
\sum_{u=1}^s(\phi^u)^2
\end{equation} is invariant under the Weyl reflections in $\G$.

Eq.(\ref{cartanquad}) is identical with the left hand side of
Eq.(\ref{gplus}). In addition, one can prove that the extremality condition
Eq.(\ref{embedding}) defines a
 a Weyl preserving embedding of ${\cal G}^{p+1}$ in $\G$  where
${\cal G}^{p+1}$ is the symmetry group resulting from dimensional reduction
of the action Eq.(\ref{oxid}) to
$(D-p-1)$ dimensions where the compactification is performed on a spatial
$p$-torus and on time. These features would provide a strong indication of
the existence of a hidden
$\G$ symmetry of the theory defined by Eq.(\ref{oxid}) or by some
generalisation there of, if the right hand side of Eq.(\ref{gplus}) would
share such group-theoretical significance. In the next section, we show,
by  construction of an explicit $\G$ invariant action, that this is indeed
the case. 

\setcounter{equation}{0}
\section{Non-linear realisation of $\G$}

This section is based on reference \cite{eh}.

\subsection{The temporal involution and the coset space ${\G}/K^{+++} $}

 We want ultimately  $\G$ to fully characterise the symmetry of the   
action $\widetilde S$ defined by Eq.(\ref{oxid}), or of some more general
theory. As an   enlargement of the {\em global} symmetry group $\cal G$ arising
in the dimensional   reduction of
$\widetilde S$,
$\G$ should also define  a {\em global} symmetry. This poses a dilemma.  
The metric and the form field strengths  in
$\widetilde S$ are genuine space-time  fields and $\widetilde S$ is invariant
under the {\em local} diffeomorphism and gauge groups. How could a global
symmetry encompass a local symmetry? The present   analysis is an attempt to
solve this dilemma by taking advantage of the infinite dimensionality
  of the algebra
$\G$. More precisely we  replace the action
$\widetilde S$ by an action $\cal S$ explicitly invariant under the global $\G$  
symmetry.
$\cal S$  contains an infinite number of objects that are tensors   under
$SL(D)$. These  comprise  a symmetric tensor $g_{\mu\nu}$,  scalars
$\phi^u$ and  ($p_I -1$)-form potentials $A_{\mu_1\mu_2 \dots  
\mu_{p_I-1}} $ which can be interpreted as the corresponding fields 
occurring in   Eq.(\ref {oxid})   taken at a {\em fixed} space time point.
Their motion in   space-time, as well as those of possible additional
fields, is expected to take place   through an infinite number of field
derivatives at this point,  encoded in other objects in
$\cal S $.

  The positive (negative) step operators in the   
$A_{D-1}$ subalgebra are, from Eq.(\ref{Kcom}), the
  $K^a{}_b$ with $b>a$ ($b<a$).  They define the level zero step  
operators of the
${\cal G}^{+++}$ adjoint representation. The positive (negative) levels  
of the adjoint representation of ${\cal G}^{+++}$ are defined as follows.
One   takes a set of
$q$ non-negative (non-positive)   integers, excluding $q$ zeros, where  
$q$ is the number of  simple roots of ${\cal G}^{+++}$ not contained in
the   gravity line. The
$q$ integers count the number of times each such root appears in the
decomposition of the adjoint representation of ${\cal G}^{+++}$ into  
irreducible representations of  $A_{D-1}$. Positive (negative) levels
contain  only   positive (negative) roots and the number of irreducible
representations of
$A_{D-1}$ at each level is finite.  All step operators of level greater
than zero may be written as irreducible tensors $ R^{\quad c_1\dots c_r}_{
d_1\dots   d_s}$ of the $A_{D-1}$ subalgebra of ${\cal G}^{+++}$, namely
\begin{equation}
\label{tensor} [K^a_{~b}, R^{\quad c_1\dots c_r}_{ d_1\dots  
d_s}]=\delta^{c_1}_b R^{\quad a\dots c_r}_{ d_1\dots
d_s}+\dots+\delta^{c_r}_b R^{\quad   {c_1}\dots a}_{ d_1\dots
d_s}-\delta_{d_1}^a R^{\quad {c_1}\dots c_r}_{b\dots   d_s}-\dots
-\delta_{d_s}^a R^{\quad {c_1}\dots c_r}_{ d_1\dots b}\, .
\end{equation}

The commutators of all positive step operators are generated by the
commutators   of
  step operators  corresponding to simple roots. At level zero these  
`simple step operators' are, from Eq.(\ref{Kcom}), the  $K^a{}_{a+1}\  
(a=1,2,\dots,D-1)$.  In general, when $s$ dilatons are present in the  
action
$\widetilde S$, the rank of
$\G$ is $D+s$. The $s$  abelian generators
$R_u$ of its subgroup $GL(D)\times U(1)^s$  have non vanishing   commutators
with the tensor step operators
$R^{a_1a_2\dots a_r}$ associated to electric or magnetic  simple roots.   In
dimensional reduction, these arise from
  $n$-form potentials where $n=p_I-1$ for an electric root and  
$n=D-p_I-1$ for a magnetic one. We can read off their commutation relation,
in the   normalisation given for the dilaton in Eq.(\ref{oxid}), namely
\cite{ehtw}
\begin{equation}
\label{dilaton} [R_u,R^{a_1a_2\dots a_r}]=
-\varepsilon\frac{a^u_I}{2}R^{a_1a_2\dots a_r}\, ,
\end{equation} where $\varepsilon=+1$ for an electric root and $-1$ for   a
magnetic one.

To switch from positive $K^a{}_b\, (b>a)$ step operators to negative   ones
it suffices to interchange  upper and lower indices.  At higher levels, the
negative   of  the tensor
$R^{\quad c_1\dots c_r}_{ d_1\dots d_s}$  is similarly a tensor $\bar
R_{c_1\dots c_r}^{\quad    d_1\dots d_s}$. The invariant scalar product for
step operators is given by
\begin{equation}
\label{normalisation}
 \langle K^b_{~a},K_{~c}^d\rangle= \delta_c^b\delta_a^d ~~ a>b ,
d>c\quad;\quad  \langle R^{\quad a_1\dots a_r}_{ b_1\dots b_s} , \bar 
R_{d_1\dots   d_r}^{\quad  c_1\dots c_s}\rangle
=\delta^{c_1}_{b_1}\dots\delta^{c_s}_{b_s}\delta^{a_1}_{d_1}\dots\delta^ 
{a_r}_{d_r}\, . 
\end{equation}

 Iterative procedures to   compute the step operators at any  level can be
devised. They build, together with   the Cartan generators
$K^a{}_a\ (a=1,2,\ldots ,D)$  and  $R_u \ (u=1,2,\dots,s)$, the full  
content of the adjoint representation  of ${\cal G}^{+++}$.

  The metric $g_{\mu\nu}$ at a fixed space time point  parametrises the  
coset
$GL(D)/SO(D-1,1)$. To construct a $\G$ invariant action $\cal S$  
containing such a tensor, we shall build   a non linear realisation of $\G$
in a coset   space
${\G}/K^{+++}$ where the subgroup $K^{+++}$ contains the Lorentz group
$SO(D-1,1)$.  We use a recursive construction based on the level  
decomposition of
$\G$. As at each level the $SO(D-1,1)$ invariance must be realised for   a
finite number of generators,  we cannot use the Chevalley involution to
build   the coset
${\G}/K^{+++}$ . Rather we shall use a `temporal'  involution  from  
which   the required non-compact generators of $K^{+++}$ can be selected.

The  `temporal'  involution $\omega$ is defined in the following way. For the  
generators of the Cartan subalgebra we take, as in the Cartan involution,
\begin{equation}
\label{cartanmap} K^a{}_a \stackrel{\omega}{\mapsto} -K^a{}_a\qquad
R_u\stackrel{\omega}{\mapsto} -R_u\, ,
\end{equation}
 while for step operators we take
\begin{equation}
\label{stepmap}  R^{\quad c_1\dots c_r}_{ d_1\dots d_s} 
\stackrel{\omega}{\mapsto} -\eta
 \bar R_{ c_1\dots c_r}^{ \quad d_1\dots d_s}\, ,
\end{equation} with
$\eta=(-1)^{n_t}$ where $n_t$ is the number of   time
indices  in $R^{\quad c_1\dots c_r}_{ d_1\dots d_s} $. All commutation  
relations in $\G$ are preserved under the mapping Eqs.(\ref{cartanmap}) and
(\ref{stepmap}),   and so is its bilinear form. This mapping constitutes an
involution that we label the temporal involution. We   define the subgroup
$ K^{+++}$ of $\G$ as the subgroup invariant under this involution. Its
generators are
\begin{equation}
\label {subgroup}  R^{\quad c_1\dots c_r}_{ d_1\dots d_s} 
 -\eta
 \bar R_{ c_1\dots c_r}^{\quad d_1\dots d_s}\, .
\end{equation}
  $ K^{+++}$ contains the Lorentz group $SO(D-1,1)$ and all generators  
with
$\eta=-1$ are non-compact .

\subsection{Non-linear realisation of $\G$ in ${\G}/K^{+++} $ }

We will follow a similar line of thought as the one  developed in reference
\cite{damourhn02} in the context of $E_8^{++}$. Consider a group element
$\cal V$  built out of Cartan and positive step operators in $\G$. It takes
the form
\begin{equation}
\label{positive} {\cal V}= \exp (\sum_{a\ge b} h_b^{~a}K^b_{~a} -  
\sum_{u=1}^s
\phi^u R_u) \exp (\sum
\frac{1}{r!s!} A^{\quad a_1\dots a_r}_{ b_1\dots b_s} R_{ a_1\dots  
a_r}^{\quad b_1\dots b_s} +\cdots)\, .
\end{equation} We have written it so that the first exponential contains
only  level zero   operators (i.e the Cartan and the level zero positive
step operators) and the second one contains the positive step operators of
level strictly greater than zero. The tensors $ h_b^{~a}, \phi^u, A^{\quad
a_1\dots a_r}_{ b_1\dots   b_s}$, bear a priori no relation with the
metric, the dilaton and the potentials  of   the $p_I$ form field strengths 
$F_{p_I}$ entering the action Eq.(\ref{oxid}). However we shall see   that a
dictionary can be established relating the  tensors which appear at low  
levels with the fields occurring in  Eq.(\ref{oxid}) at a {\em fixed}  
space-time point. For higher levels the dictionary between group parameters
and space-time   fields should arise from the analysis of the   dynamics
encoded in the
$\G$ invariant action
$\cal S$ below.

A differential motion in the coset ${\G}/K^{+++}$ can be constructed   from
Eq.(\ref{positive}). Define
\begin{equation}
\label {sym} dv= d{\cal V} {\cal V}^{-1}\qquad d\tilde v= -\omega dv 
\qquad;\qquad dv_{sym}=\frac{1}{2} (dv+d\tilde v)\, .
\end{equation}  As $dv$ and $d\tilde v$   are differentials in the Lie
algebra,  $dv_{sym}$ contains only the Cartan   generators and the
combinations of step operators 
$ R^{\quad c_1\dots c_r}_{ d_1\dots d_s} +
\eta \bar R_{ c_1\dots c_r}^{\quad d_1\dots d_s}$.

To construct the action $\cal S$ we wish  to map  a manifold
$\cal M$ into
$\G$. We do not want to take for
$\cal M$ a space-time manifold, as this might require the explicit  
introduction of local symmetries which we hope to be hidden in the infinite
algebra   of
$\G$.  We shall take for $\cal M$  a one-dimensional world-line in
$\xi$-space, i.e.
$dv_{sym}= dv_{sym}(\xi)= d{\cal V}(\xi) {\cal V}^{-1}(\xi)$, where ${\cal V(\xi)}$ are the group
parameters appearing   in Eq.(\ref{positive}) that are now fields dependent on
the  variable  
$\xi$. Here no
connection is imposed a priori between  
$\xi$-space and space-time.

A reparametrisation invariant action is then
\begin{equation}
\label{actionG} {\cal S}=\int d\xi 
\frac{1}{n(\xi)}\langle(\frac{dv_{sym}(\xi)}{d\xi})^2\rangle\, ,\end{equation}
 where
$n(\xi)$ is an arbitrary lapse function ensuring reparametrisation  
invariance on the world-line. The `trace' $\langle~\rangle$ is given by
Eqs.(\ref{cartanscalar}) and (\ref{normalisation}) . It ensures the
invariance of the non-linear action $\cal S$ defined on the coset space
${\G}/K^{+++} $ under global $\G$ transformations.

We now compute the level zero of the action Eq.(\ref{actionG}), that is  
the terms generated by  $K^b_{~a}~(a\ge b)$ and $R_u$ in $\cal V$.  
From Eqs.(\ref{positive}), (\ref{sym}) and (\ref{Kcom}), one obtains the
contribution   of the level zero to $v_{sym}(\xi)$,
\begin{equation}
\label{v0} \frac{dv_{sym}^0(\xi)}{d\xi}=-\frac{1}{2}\sum_{a\ge b} [e^  h
(\frac{de^{- h}}{d\xi})]_b^{~a} (K^b_{~a}+
\eta K^{a}_{~b}) -\sum_{u=1}^s
\frac{d\phi^u}{d\xi}  R_u\, ,
\end{equation} where $ h$ is triangular matrix with elements   
$h_b^{~a}$. We now evaluate $\langle (dv_{sym}^0(\xi)/d\xi)^2\rangle$
using Eqs.(\ref{cartanscalar}) and (\ref{normalisation}). The $\eta$-symbols
defining the temporal involution   allow the raising or lowering of the
$a,b$ indices of the  
$\xi$-fields multiplying the negative step operator in
$\langle dv_{sym}^0/d\xi, dv_{sym}^0/d\xi\rangle$ with the
Minkowskian metric
$\eta_{ab}$ . This ensures that this   expression is a Lorentz scalar. The
Lorentz invariant action at level   zero, ${\cal S}^0$, is
\begin{eqnarray}
\label{sym2} {\cal S}^{(0)}&=&\int d\xi
\frac{1}{n(\xi)}\langle(\frac{dv_{sym}^0(\xi)}{d\xi})^2\rangle\, ,
\nonumber\\ &=&\frac{1}{2}\int d\xi
\frac{1}{n(\xi)}\left\{  [e^  h (\frac{de^{-  h}}{d\xi})]_b^{~a} [e^  h  
(\frac{de^{- h}}{d\xi})]^{T~b}_a +[e^  h (\frac{de^{-  h}}{d\xi})]_b^{~a}
[e^  h   (\frac{de^{-
h}}{d\xi})]_a^{~b}\right.\nonumber\\&&\qquad\qquad\qquad\left.- ([e^  h
(\frac{de^{-  h}}{d\xi})]_a^{~a})^2 +
\sum_{u=1}^s (\frac{d\phi^u}{d\xi})^2\right\}  ,
\end{eqnarray} where the summation is performed over Lorentz indices.   Note
that the lower indices of
$e^{-h}$ and the upper indices of $e^h$ cannot be lowered or raised by   the
Lorentz metric. To avoid confusion we label these indices with greek
letters,   namely we define `vielbein'
\begin{equation}
\label{vielbein}
  e_\mu^{~a}=(e^{-h})_\mu^{~a}\quad e^{~\mu}_b=(e^h)^{~\mu}_b\quad;\quad
g_{\mu\nu} =e_\mu^{~a}e_\nu^{~b}\eta_{ab}\, .
\end{equation} Although we have not  yet introduced a space-time, we   shall
name the $a$ indices  flat  and the $\mu$ indices  curved. As a result   of
the temporal involution and of the scalar product $\langle~\rangle$\ in  
$\G$, the flat-index tensors have been endowed with a Lorentz metric while  
curved-index tensors define a metric in $GL(D)/SO(D-1,1)$. Hence, for any
$\xi$, we   are allowed to identify
$g_{\mu\nu}(\xi)$ in Eq.(\ref{vielbein}) as the metric tensor in $\widetilde S$,  
Eq.(\ref{oxid}), at a fixed space-time point.

Using Eq.(\ref{vielbein}), one can  rewrite the action Eq.(\ref{sym2})   as
\begin{eqnarray}
\label{action0} {\cal S}^{(0)}&=&\frac{1}{2}\int d\xi
\frac{1}{n(\xi)}\left[\frac{1}{2}(g^{\mu\nu}g^{\sigma\tau}- 
\frac{1}{2}g^{\mu\sigma}g^{\nu\tau})\frac{dg_{\mu\sigma}}{d\xi}
\frac{dg_{\nu\tau}}{d\xi}+\sum_{u=1}^s
\frac{d\phi^u}{d\xi}\frac{d\phi^u}{d\xi}\right].
\end{eqnarray} 

At higher levels, the tensors multiplying the step   operators couple
nonlinearly to the level zero objects and  between themselves.   The
coupling to the metric and to $\phi^u$ can be  formally written down   for
all levels, but the self-coupling of the $A^{\quad a_1\dots a_r}_{ b_1\dots
b_s}  
$depend specifically on the group
$\cal G$.

  Consider a general
$A_{D-1}$ tensor $A_{a_1\dots a_r}^{\quad b_1\dots b_s}$ parametrising a
normalised step operator
$R^{\quad a_1\dots a_r}_{ b_1\dots b_s}$. The commutation relations of  
$R^{\quad a_1\dots a_r}_{ b_1\dots b_s}$ with the $K^b_{~a}$ are given by
the   tensor transformations as in Eq.(\ref{Kcom}) and (\ref{tensor}) and
those with  
$R_u$ have the form
\begin{equation}
\label{gendilaton} [R_u,R^{\quad a_1\dots a_r}_{ b_1\dots b_s}]=  
\lambda_u\, R^{\quad a_1\dots a_r}_{ b_1\dots b_s}\, .
\end{equation} Here
$\lambda_u=\sum\lambda_{u,i}$ where the $\lambda_{u,i}$ are the scale
parameters of the simple step operators entering the multiple   commutators
defining $R^{\quad a_1\dots a_r}_{ b_1\dots b_s}$. This property   follows 
from the Jacobi identity. Identifying for simple step operators
$\lambda_{u,i}$   with
$-\varepsilon a^u_I/2$ in Eq.(\ref{dilaton}) we may  identify for any  
$\xi$ the
$\phi^u(\xi)$ in $\cal S$, Eq.(\ref{actionG}), with the dilatons fields   in
$\widetilde S$, Eq.(\ref{oxid}),  at a fixed space time point. The particular
$A_{a_1\dots a_r}^{\quad b_1\dots b_s}(\xi)$ multiplying the step  
operators belonging to the subgroup $\cal G$ can be similarly identified to
the corresponding potential forms in $\widetilde S$ along with their duals.

It is  straightforward to compute the contribution
$dv^{(A)}$ to
$dv$ of a given tensor when commutators of the $R^{\quad a_1\dots a_r}_{  
b_1\dots b_s}$ between themselves are disregarded. On gets
\begin{equation}
\label{vform} dv^{(A)}=\frac{1}{r!s!}dA_{\mu_1\dots \mu_r}^{\quad  
\nu_1\dots
\nu_s} \, \exp (-\sum_{u=1}^s
\lambda^u \phi^u)\, e^{~\mu_1}_{{a}_1}\dots  
e^{~\mu_r}_{{a}_r}e_{\nu_1}^{~{b}_1}
\dots e_{\nu_s}^{~{b}_s}\, R^{\quad a_1\dots a_r}_{ b_1\dots b_s}\,  .
\end{equation} The contribution ${\cal S}^{(A)}_0$ of $v^{(A)}$ to the  
action
${\cal S}$ is computed as previously and one gets
\begin{equation}
\label{actionA} {\cal S}^{(A)}_0=\frac{1}{2}\int d\xi
\frac{1}{n(\xi)}\left[\frac{1}{r!s!}\exp (-\sum_{u=1}^s 2\lambda^u  
\phi^u)
\frac{dA_{\mu_1\dots \mu_r}^{\quad \nu_1\dots
\nu_s}}{d\xi} g^{\mu_1{\mu}^\prime_1}...\,
g^{\mu_r{\mu}^\prime_r}g_{\nu_1{\nu}^\prime_1}...\,  
g_{\nu_s{\nu}^\prime_s}
\frac{dA_{{\mu}^\prime_1\dots {\mu}^\prime_r}^{\quad {\nu}^\prime_1\dots
{\nu}^\prime_s}}{d\xi}\right].
\end{equation} The full action can only be approached in a recursive   way.
In
${\cal S}^{(A)}_0$, one must replace derivatives by non linear
generalisations   to take into account the non vanishing commutators
between tensor step   operators. We represent such terms by `covariant'
derivatives symbol
$D/D\xi$ \cite{damourhn02,eh}. There evaluation is group dependent.  Formally
the full action
$\cal S$ is
\begin{equation}
\label{full} {\cal S}={\cal S}^{(0)}+\sum_A{\cal S}^{(A)}\, ,
\end{equation}
$${\cal S}^{(A)}=\frac{1}{2}\int d\xi
\frac{1}{n(\xi)}\left[\frac{1}{r!s!}\exp (-\sum_{u=1}^s 2\lambda^u  
\phi^u)
\frac{DA_{\mu_1\dots \mu_r}^{\quad \nu_1\dots
\nu_s}}{d\xi} g^{\mu_1{\mu}^\prime_1}...\,
g^{\mu_r{\mu}^\prime_r}g_{\nu_1{\nu}^\prime_1}...\,   
g_{\nu_s{\nu}^\prime_s}
\frac{DA_{{\mu}^\prime_1\dots {\mu}^\prime_r}^{\quad {\nu}^\prime_1\dots
{\nu}^\prime_s}}{d\xi}\right] $$ where the sum on $A$ is a summation   over
all tensors appearing at all positive levels in the decomposition of $\G$  
into irreducible representations of $A_{D-1}$.

One may expand $\cal S$ given in Eq.(\ref {full}) in power of fields  
parametrising the positive step operators. Up to quadratic terms, the result
${\cal   S}^{(Q)}$ is obtained by retaining in
$v(\xi)$ terms independent or linear in these fields. Define the   one-forms
$[dA]$ and the moduli $p^{(a)}$ (or $p^{(\mu)}$) as in Section 4.3 by
\begin{eqnarray}
\label{flatform}
\exp (-\sum_{u=1}^s
\lambda^u \phi^u)\,
\hat e^{~\mu_1}_{{a}_1}\dots\hat e^{~\mu_r}_{{a}_r}\hat   e_{\nu_1}^{~{b}_1}
\dots\hat e_{\nu_s}^{~{b}_s}\,\frac{dA_{\mu_1\dots \mu_r}^{\quad  
\nu_1\dots
\nu_s}}{d\xi}&\stackrel{def}{=}&\frac{ [dA]_{a_1\dots a_r}^{\quad   b_1\dots
b_s}}{ d\xi}\, ,\\
\label{flatmetric}
\hat e^{~\mu}_b\frac{d\hat e_{\mu}^{~a}}{d\xi}&\stackrel{def}{=}&\frac   {d
p^{(a)}}{d\xi}\, ,
\end{eqnarray} where   $\hat e$ means that only the diagonal vielbein   are
kept. We get
\begin{eqnarray}
\label{linear} {\cal S}^{(Q)}&=&\int d\xi\frac{1}{n(\xi)}  
\left[\sum_{a=1}^D (\frac{dp^{(a)}}{d\xi})^2-\frac{1}{2}(\sum_{a=1}^D  
\frac{dp^{(a)}}{d\xi})^2 +\frac{1}{2}\sum_{u=1}^s
(\frac{d\phi^u}{d\xi})^2\right.\nonumber \\
&&\left.+\frac{1}{2}(e^{~\mu}_b\frac{d e_\mu^{~a}}{d\xi}\frac{d  
e_{a\nu}}{d\xi}
  e^{\nu b})^{(1)}+\frac{1}{2}\frac{1}{r!s!}\sum_A\frac{[dA]_{a_1\dots  
a_r}^{\quad b_1\dots b_s}}{ d\xi}\, \frac{[dA]^{\quad a_1\dots a_r}_{
b_1\dots   b_s}}{ d\xi}\right]  ,
\end{eqnarray} where the superscript {\footnotesize (1)} in the vielbein
term indicates that   only terms quadratic in $\, h_b{}^a\, (a>b)$ are
kept.  In the next section we   shall produce solutions of the  action
Eq.(\ref{linear})  which are {\em exact}   solutions of the full action
Eq.(\ref {full}).

\setcounter{equation}{0}
\section{Extremal branes from the $\G$ invariant actions}

We shall look for solutions of the equations of motion derived from   
$\cal S$ and containing only one
$A(\xi)$-field,  or one non-diagonal $h(\xi)$-field,  with {\em given
indices} \cite{eh}.   Such truncation is consistent with all the equations of
motion and hence we may disregard all non-linearity in the step operators. 
Therefore it suffices   for  obtaining such solutions to replace the
action  $\cal S$ by its     simplified version Eq.(\ref{linear}).

We shall as in Eqs.(\ref{electric}) and (\ref{magnetic}), consider $A$   to
be an antisymmetric tensor with a time index  $\tau$ and  
$r$ space indices coupled to a step operator of the  $\cal G$ subalgebra.
The   equation of motions are

\begin{itemize}

\item[]{\it a) The lapse constraint}.

Eq.(\ref{linear}), taking Eqs.(\ref{flatform}) and (\ref{flatmetric}) into
account, reads
\begin{equation}
\label{lapse}
\sum_{\alpha=1}^D  
(\frac{dp^{(\alpha)}}{d\xi})^2-\frac{1}{2}(\sum_{\alpha=1}^D
\frac{dp^{(\alpha)}}{d\xi})^2 +\frac{1}{2} (\frac{d\phi}{d\xi})^2   
-\frac{1}{2} \exp [
\varepsilon a \phi -2p^{(\tau)}-2\sum_{\lambda=\lambda_1}^{\lambda_r}
p^{(\lambda)})] (\frac{dA_{\tau\lambda_1\dots \lambda_r}}{ d\xi})^2=0
\end{equation} Here we have taken one dilaton with scaling $\lambda
=-\varepsilon a/2$ in accordance with Eq.(\ref{dilaton}). Note that   this
relation is valid wether or not the  magnetic root is simple, as seen in
dimensional reduction. A crucial feature   of this equation is the minus
sign in front of the   exponential. Its origin can be traced back to the
temporal  involution defining our coset space, hence to Lorentz invariance,
because {\em both} magnetic  and electric potentials have a   time index.

\item[]{\it b)  The equation of motion for $A$}.

We take the lapse $n(\xi)=1$. One gets
\begin{equation}
\label {potential}
\frac{d}{ d\xi} \left( \, \exp [
\varepsilon a \phi -2p^{(\tau)}-2\sum_{\lambda=\lambda_1}^{\lambda_r}
p^{(\lambda)}]\frac{dA_{\tau\lambda_1\dots
\lambda_r}}{ d\xi} \, \right) =0\, .
\end{equation}

\item[]{\it c) The dilaton equation of motion}.
\begin{equation}
\label{dilat} -\frac{d^2\phi}{ d\xi^2}-\frac{1}{2}\varepsilon a\exp [
\varepsilon a \phi -2p^{(\tau)}-2\sum_{\lambda=\lambda_1}^{\lambda_r}
p^{(\lambda)}](\frac{dA_{\tau\lambda_1\dots
\lambda_r}}{ d\xi})^2=0\, .
\end{equation}

\item[]{\it d) The vielbein equations of motion}.
\hskip -1cm\begin{equation}
\label{vielbein1} -2\frac{d^2p^{(\alpha)}}{ d\xi^2} +\sum_{\beta=1}^D
\frac{d^2p^{(\beta)}}{ d\xi^2}=0\qquad
\alpha\neq \tau ,\lambda_i~~ (i=1,2\dots r)\, ,
\end{equation}
\begin{equation}
\label{vielbein2} -2\frac{d^2p^{(\alpha)}}{ d\xi^2} +\sum_{\beta=1}^D
\frac{d^2p^{(\beta)}}{ d\xi^2}+\exp [
\varepsilon a \phi -2p^{(\tau)}-2\sum_{\lambda=\lambda_1}^{\lambda_r}
p^{(\lambda)}](\frac{dA_{\tau\lambda_1\dots
\lambda_r}}{ d\xi})^2=0\quad\alpha= \tau ,\lambda_i\, .
\end{equation}
\end {itemize} We take as anz\"atze the solutions of the extremal brane  
problem but with $H_A$ an  unknown function of $H(\xi)$. Namely we pose
\begin{eqnarray}
\label{xx} A_{\tau\lambda_1\dots
\lambda_r}&=&\epsilon_{\tau
\lambda_1\dots\lambda_r}[\frac{2(D-2)}{\Delta}]^{1/2}H^{-1}(\xi)\, ,\\
\label{yy}
  p^{(\tau)}=p^{(\lambda_i)}= - \frac{D-r-3}{\Delta}  \ln H(\xi)~   &;& ~
p^{(\alpha)}=\frac{r+1}{\Delta} \ln H(\xi)\quad\alpha\neq \tau  
,\lambda_i\, .\\
\label{zz}
\phi &=& \frac{D-2}{\Delta}\varepsilon a \ln H(\xi) \, .
\end{eqnarray} From these equations and from Eq.(\ref{delta}) we see   that
$\varepsilon a
\phi -2p^{(\tau)}-2\sum_{\lambda=\lambda_1}^{\lambda_r}   p^{(\lambda)}=2\ln
H(\xi)$.  It then follows that  the  equation of motion for $A$,  
Eq.(\ref{potential}), reduces to, using Eq.(\ref{xx}),
\begin{equation}
\label{basic}
\frac{d^2H(\xi)}{ d\xi^2}=0\, .
\end{equation} Given this result, it is straightforward to verify that the 
anz\"atze   Eqs.(\ref{xx}), (\ref{yy}) and (\ref{zz})  satisfy the dilaton
and the vielbein equations of motions. The lapse constraint takes the form
\begin{equation}
\label{xiextremal}
\sum_{\alpha=1}^D ( d p^{(\alpha)})^2 -\frac{1}{2}(\sum_{\alpha=1}^D
dp^{(\alpha)})^2 +
\frac{1}{2} (d\phi)^2- \frac{D-2}{\Delta}(d\ln H)^2=0\, .
\end{equation} where the differentials are taken in $\xi$-space. It has  
therefore exactly the same form  in
$\xi$-space as  Eq.(\ref{gplus}) has in space-time.  The relations  
Eqs.(\ref{xx}), (\ref{yy}) and (\ref{zz}), together with Eqs.(\ref{basic})
and the   lapse constraint Eq.(\ref{xiextremal}) fully describe an exact
solution of the full
$\G$ invariant action $\cal S$ defined recursively by Eq.(\ref{full}).   We
now discuss the significance of this result.

The  Eqs.(\ref{xx}), (\ref{yy}) and (\ref{zz})  characterise completely  
the algebraic structure of the extremal brane solution but do not yield
its   harmonic character in space-time. As the functions
$A_{\tau\lambda_1\dots
\lambda_r}(\xi),\, p^{(\tau)}(\xi),\,  p^{(\lambda_i)}(\xi) $ and $  
\phi(\xi)$ were interpreted in the action $\cal S$ as functions at a fixed
space-time   point of the independent variable
$\xi$, this is a consistent result.  The solution $H = a+ b\xi$ of  
Eq.($\ref{basic}$) would then describe a motion in the space of solutions,
for instance of   branes with different charges.  However the fact that we
have  exact solutions of   the action
$\cal S$ with the correct algebraic structure of the extremal branes,  
means that these solutions  are only  indirectly related to the
corresponding     space-time solution. One expects that the information
contained in this solution,   which is of course  not contained in a
trivial  constant space-time solution of the   Einstein equation, is the
required information to build coupled equations to   higher space-time
derivatives encoded in  higher level representations, which   would then be
directly related to space-time solutions.

There are indications that this may indeed be the case but it is unclear
whether it is sufficient to consider  the adjoint representation of $\G$
\cite{eh} or if one has  to include other representations
\cite{west04,kw}. 

The solution Eqs.(\ref{xx}), (\ref{yy}) and (\ref{zz}) satisfy, in  
$\xi$-space, the relation Eq.(\ref{embedding}). This relation define an
embedding of a   subgroup
${\cal G}^{p+1}$ of $\G$ acting on the $p$  compact space dimensions    in
which the branes live and on the time dimension \cite{ehw}.  We shall
consider the subgroup  
${\cal G}^p$ of
${\cal G}^{p+1}$ which acts on the space dimensions only and we take  
$p\le D-4$ so that
${\cal G}^{p+1}$ is a Lie group. This group is conjugate by a Weyl  
reflection in  $\G$ of the group ${\cal G}^{\prime\, p+1}$ obtained by
deleting the first  
$D-p-1$ nodes of the gravity line \cite{ehw} and hence ${\cal G}^p$ is
conjugate to its subgroup  
${\cal G}^{\prime\, p}$ characterising the usual dimensional reduction of  
Eq.(\ref{oxid}) to $D-p$ dimensions.

  We shall consider  the  transformations mapping {\em one}  root to  
another   root, thereby  generating solutions of the same `family' as the
extremal   solution just described.  These transformations  include the 
Weyl group $W({\G}) $
  of $\G$. We shall   examine   Weyl transforms of the extremal brane  
solution characterised by one positive step operator which  send the
positive   root into a positive root. Such transformations leave invariant
not only
$\cal S$ but also preserves their quadratic truncation   Eq.(\ref{linear}).
Hence Eq.(\ref{xiextremal})  is invariant under the  Weyl group of ${\cal
G}^{p+1}$. The restriction   to  the Weyl group of ${\cal G}^p$ selects 
transformed fields with    one time index.

Thus $W({\cal G}^p) $ leaves invariant   the quadratic form
\begin{equation}
\label{lapsegen}
\sum_{a=1}^D (dp^{(a)})^2+\frac{1}{2} [-(\sum_{i=1}^D dp^{(a)})^2  
+\sum_{u=1}^s (d\phi^u)^2 +(e^{~\mu}_b d e_\mu^{~a}d e_{a\nu}
  e^{\nu b})^{(1)}+\frac{1}{r!s!}\sum_A \,  [dA]_{a_1\dots a_r}^{~  
b_1\dots b_s}\, [dA]^{~ a_1\dots a_r}_{ b_1\dots b_s} ] ,
\end{equation} and the  embedding relation Eq.(\ref{embedding}) in  
$\xi$-space. It acts on $A$-fields, or non-diagonal vielbein, containing
one time index.   The sum of the first three terms is the invariant metric
of $\G$ restricted to its   Cartan subgroup. Together with the embedding
relation Eq.(\ref{embedding})   they are left invariant under the Weyl
group of ${\cal G}^{p+1}$. We now see the group theoretical relevance of the
right hand side of Eq.(\ref{gplus}) or better the last term of 
Eq.(\ref{xiextremal}) in
$\xi$-space. The Weyl transformations of the step operators and
the additional terms in Eq.(\ref{lapsegen}) guarantee  indeed the invariance of 
Eq.(\ref{xiextremal})   under 
$W({\cal G}^p) $.   For all $\G$, the    Weyl
transformations generate  new   solutions from one extremal brane solution. We
stress again that in the present   approach both electric and magnetic
branes are described `electrically'.

For M-theory, this yield the well-known duality symmetries of M-theory,
including the duality transformations of branes, KK waves and KK monopoles
(Taub-NUT spaces).   Such  transformations are however not a privilege of
M-theory and occur   in {\em all} $\G$ invariant actions. This is
exemplified below, taking for   definiteness the action $\cal S$ for the
group $E_7^{+++}$ which is related to the   action $\widetilde S$ with gravity
coupled to a 4- and a 2- form field strength in 9 space-time dimensions.
The Dynkin diagram of  
$E_7^{+++}$ is depicted in Fig.3, which exhibits the two simple electric
roots
$(10)$ and $(9)$ corresponding  respectively to the step operators
$R^{7\,8\,9}$ and $R^9$ which couple  to the electric potentials
$A_{7\,8\,9}$ and
$A_9$.

We take as input the electric extremal 2-brane ${\bf e}_{(8,9)}$ in the  
directions
$(8,9)$ associated with the 4-form field strength whose corresponding  
potential is
$A_{1\,8\,9}$ and submit it to the non trivial Weyl reflection $W_{10}$  
associated with the electric root
$(10)$ of Fig.3. We display below, both for ${\bf e}_{(8,9)}$ and its  
transform,  the moduli, i.e. the vielbein components
$p^{(a)}$ and the the dilaton value $\phi$, of the brane solution  
Eqs.(\ref{yy}) and (\ref{zz}) as a ten-dimensional vector where the last
component is the   dilaton.  We also indicate the transform of the step
operator $R^{1\,8\,9}$ under   the Weyl transformation.   We obtain

\begin{eqnarray}
\label {e89} (-4, 3,3,3,3,3,3, -4,-4;2\sqrt 7)\, \frac{\ln H(\xi)}{14}&  
{\bf e}_{(8,9)} &R^{1\,8\,9}\\ &\qquad\downarrow W_{10}\quad&\nonumber\\
\label{k17} (-7, 0,0,0,0,0,7, 0,0;0)\, \frac{\ln H(\xi)}{14}& {\bf  
kk_e}_{\, (7)} &K^1_{~7}
\end{eqnarray} The transformation of the 2-brane is  reminiscent of a double
T-duality  in   M-theory.

We now move the electric brane through Weyl reflections associated with  
roots of the gravity line to ${\bf e}_{(5 ,9)}$ and submit it to the Weyl  
reflection $W_{10}$. We now find that the brane ${\bf e}_{(5 ,9)}$ is 
invariant but   moving it to the position
${\bf e}_{(5 ,6)}$, we get
\begin{eqnarray}
\label {e56} (-4, 3,3,3,-4,-4,3,3,3;2\sqrt 7)\, \frac{\ln H(\xi)}{14}&  
{\bf e}_{(5,6)} &R^{1\,5\,6}\\ &\qquad\downarrow W_{10}\quad&\nonumber\\
\label{m56789} (-1, 6,6,6,-1,-1,-1,-1,-1;4\sqrt 7)\, \frac{\ln  
H(\xi)}{14}& {\bf m}_{(5,6,7,8,9)}  &R^{1\,5\,6\, 7\,8\,9}
\end{eqnarray} This is a magnetic 5-brane in the directions  
$(5,6,7,8,9)$ associated to the  2-form field strength~!   It is expressed
in terms   of its dual potential $A_{1\,5\, 6\,7\,8\,9}$. Submit instead
${\bf e}_{(5 ,9)}$ to   to the Weyl reflection $W_{9}$ associated with the
electric root $(9)$ of Fig.3.   The 2-brane ${\bf e}_{(5   ,9)}$  is again 
invariant, but moving it to to the position
${\bf e}_{(5 ,6)}$, we now get
\begin{eqnarray} (-4, 3,3,3,-4,-4,3,3,3;2\sqrt 7)\, \frac{\ln   H(\xi)}{14}&
{\bf e}_{(5,6)} &R^{1\,5\,6}\nonumber \\ &\qquad\downarrow  
W_{9}\quad&\nonumber\\
\label{m569} (-3, 4,4,4,-3,-3,4,4,-3;-2\sqrt 7)\, \frac{\ln   H(\xi)}{14}&
{\bf m}_{(5,6,9)}  &R^{1\,5\,6\, 9}
\end{eqnarray} This is a magnetic 3-brane in the directions $(5,6,9)$  
associated to the  4-form field strength, expressed in terms of its dual
potential  
$A_{1\,5\, 6\,9}$.

Finally, let us submit the magnetic 5-brane $ {\bf m}_{(5,6,7,8,9)}$  
obtained in Eq.(\ref{m56789}) to the Weyl reflection  $W_{9}$. One obtains
\begin{eqnarray} (-1, 6,6,6,-1,-1,-1,-1,-1;4\sqrt 7)\, \frac{\ln  
H(\xi)}{14}& {\bf m}_{(5,6,7,8,9)}  &R^{1\,5\,6\, 7\,8\,9}\nonumber \\
&\qquad\downarrow W_{9}\quad&\nonumber\\
\label{h2349} (0, 7,7,7,0,0,0,0,-7;0)\, \frac{\ln H(\xi)}{14}&  {\bf  
kk_m}_{\, (2,3,4;9)} &R^{1\,5\,6\,7\,8\,9,\,9}
\end{eqnarray} Eq.(\ref{h2349}) describes, as in M-theory, a purely
gravitational     configuration, namely a KK-monopole  with transverse  
directions (2,3,4) and Taub-NUT direction (9) in terms of a dual gravity
tensor
$h_{1\,5\,6\,7\,8\,9,\,9}$.

It is also possible  to generate solutions not   contained, at least
explicitly, in the group ${\cal G}^p$. These are very interesting  
solutions as they may test the significance of genuine Kac-Moody extensions
of the Lie   groups. Such analysis is outside the scope of the present
work where we test only   solutions which can straightforwardly be mapped 
to space time solutions of the effective   actions Eq.(\ref{oxid}).

The above example illustrate the analogy of M-theory duality  
transformations with similar transformations in all `M-theories' defined by
all $\G$.  One   may indeed carry the same analysis for all $\G$  and exhibit for
each of them the   `duality' transformations of the branes. As in M-theory, these
dualities are   symmetries in non-compact space-time.  This is because
${\cal G}^{ p+1}$ is, as ${\cal G}^{\prime\, p+1}$, the Lie group  
symmetry  of the action Eq.(\ref{oxid}) dimensionally reduced to three
dimensions (for  
$p=D-4$). They differ because while the latter reduction leaves a
Lorentzian   non-compact space-time, the former leads to a Euclidean
space-time by   compactifying time. The group ${\cal G}^p$ of
transformations on $\xi$-space  discussed   above, is thus in one to one
correspondence with the group ${\cal G}^p$ of space-time transformations
when time is decompactified. In particular, the   functions
$H(\xi)$ can thus be mapped into  harmonic functions $H(\{x^\nu\})$.  
However as pointed out in the previous section, more work is needed to
relate   directly
$H(\xi)$ to
$H(\{x^\nu\})$, and solutions in
$\xi$-space to solutions in space-time for all $\G$, through translation  
operators hopefully induced by group generators. 

\section*{Acknowledgments}

This work was supported in part  by the NATO grant PST.CLG.979008,
  by the ``Actions de Recherche Concert\'ees'' of the ``Direction de la
Recherche Scientifique - Communaut\'e Fran\c caise de Belgique, by a
``P\^ole d'Attraction Interuniversitaire'' (Belgium), by IISN-Belgium
(convention 4.4505.86), by Proyectos FONDECYT 1020629, 1020832 and 7020832
(Chile) and by the European Commission RTN programme HPRN-CT00131, in which
F.~E. and L.~H. are associated to the Katholieke Universiteit te Leuven
(Belgium).

\end{document}